\documentstyle[aps,prbbib,twocolumn,epsf]{revtex}

\begin{document}
\draft
\title{Quantum spin field effect transistor}

\author{Baigeng Wang$^1$, Jian Wang$^{1,2}$, and Hong Guo$^3$}
\address{1. Department of Physics, The University of Hong Kong, 
Pokfulam Road, Hong Kong, China\\
2. Institute of Solid State Physics, Chinese Academy of Sciences,
Hefei, Anhui, China\\
3. Department of Physics, McGill University, Montreal, Quebec, 
Canada H3A 2T8}
\maketitle

\begin{abstract}
We propose, theoretically, a new type of quantum field effect transistor 
that operates purely on the flow of spin current in the absence of charge 
current. This spin field effect transistor (SFET) is constructed without
{\it any} magnetic material, but with the help of spin flip mechanism 
provided by a rotating external magnetic field of uniform strength. The SFET
generates a {\it constant} instantaneous spin current that is sensitively
controllable by a gate voltage as well as by the frequency and strength of 
the rotating field. The characteristics of a Carbon nanotube based SFET is
provided as an example.
\end{abstract}


Si-based field effect transistors (FET) have played a pivotal role in the
the technology that drives the microelectronics revolution. It has however 
been projected that Si technology is rapidly approaching its limit of 
miniaturation\cite{wada}, and various new and exciting ideas of 
nano-electronics have been proposed and pursued. One of the most important
possibilities of nano-electronics is the hope of using spin---in addition to
charge, for nonlinear electronic device applications\cite{review}. So far, 
progress has been achieved in certain areas of spintronics such as device 
applications based on the giant magnetoresistive effect\cite{baibich}, the 
understanding of material properties of magnetic semiconductors\cite{ohno}, 
the improvements of spin injection across a magnetic---nonmagnetic 
interface\cite{injection}, and optical manipulation of spin degrees of freedom 
in nanostructures\cite{awschalom}. On the other hand, despite the fact that 
it is already more than ten years since the proposal\cite{das} of FET 
operation on spin-polarized charge current, the spin-FET (SFET) has been an 
elusive system up to now. The overwhelming majority of actual spintronics 
devices and proposals up to now are hybrid systems which involve both magnetic 
and non-magnetic materials\cite{review}. Due to differences in chemical 
bonding and structural property of them, these hybrid materials have proven 
to be rather challenging to use. This, together with several other
physical factors related to spin transport, has limited the rapid
development of practical nonlinear spintronic devices such as the SFET.

In this paper, we take a different direction by theoretically examining the
possibility of SFET operation {\it without} involving magnetic materials and
we exploit such a novel SFET which operates purely on {\it spin current}.
This SFET turns out to be realizable---as we predict, in coherent 
nanostructures (such as a quantum dot, a quantum well, or a Carbon nanotube), 
in the presence of a rotating external magnetic field of uniform strength. 
Importantly, the rotating field induces a {\it time-independent} ({\it i.e.}
a DC) spin current, and at the same time it generates no charge current (see 
below).  The magnitude of the spin current is critically tunable by a gate 
voltage which shifts the electronic levels of the non-magnetic nanostructure 
so that SFET operation is achieved. The physical principle of our SFET is due 
to spin flip mechanism provided by the field, but is ultimately connected to 
the quantum physics of Berry's phase\cite{berry}. Because no magnetic 
material is involved in our SFET, any problem that relates to spin injection 
across a magnetic-nonmagnetic interface is bypassed. Moreover, because there 
is no charge current involved, our SFET will be less affected by problems of 
heat dissipation. Since many non-magnetic nanostructures, such as a Carbon 
nanotube, have long spin coherent lengths\cite{tsukagoshi}, our proposed 
quantum SFET should be experimentally realizable. To provide a concrete 
numerical example, we predict the transport characteristics of an 
all-nanotube based SFET.  

Consider a three-probe non-magnetic device shown in the inset of Fig.1 
which consists of a scattering region Ohmic-contacted by two probes while 
the third probe is a metallic gate capacitively coupled to the scattering 
region. Here we used a section of an armchair carbon nanotube as the scattering 
region, but in general it can be a quantum dot, a quantum well, or other
mesoscopic conductors. The system can be 2d or 3d. In this work, we 
consider the following Hamiltonian of this device ($\hbar=1$):
\begin{eqnarray}
H &=&\sum_{k,\sigma ,\alpha =L,R}\epsilon _{k}C_{k\alpha \sigma
}^{+}C_{k\alpha \sigma }+\sum_{\sigma }[\epsilon +\sigma B_{0}\cos \theta
]d_{\sigma }^{+}d_{\sigma } \nonumber \\
&&+H'(t)  
+\sum_{k,\sigma ,\alpha =L,R}[T_{k\alpha }C_{k\alpha \sigma }^{+}d_{\sigma
}+c.c.]
\label{eq1}
\end{eqnarray}
where $H'(t)$ is the off diagonal part (in spin space) of the Hamiltonian,
\begin{equation}
H^{\prime }(t)= \gamma ~ \lbrack \exp (-i\omega t)d_{\uparrow
}^{+}d_{\downarrow }+\exp (i\omega t)d_{\downarrow }^{+}d_{\uparrow }] \ ,
\label{hp}
\end{equation}
with $\gamma=B_o \sin\theta$.  In the Hamiltonian (\ref{eq1}), the first term 
stands for the Hamiltonian of noninteracting electrons in the leads with 
$C_{k\alpha \sigma}^\dagger$ the creation operators in lead $\alpha$. Note 
that we have set the same chemical potential for both leads because, as we 
will see below, a rotating magnetic field will ``pump'' out a DC spin current 
without needing a bias voltage. Quantum parametric pumping of {\it charge 
current} has been well analyzed before\cite{brouwer}, and here we 
demonstrate that puming of a spin current will lead to SFET operation. The 
second term and $H'(t)$ correspond to the Hamiltonian of the scattering region 
which is subjected to a time-dependent (rotating) magnetic field with uniform 
strength, ${\bf B}(t)=B_o~ [\sin\theta\cos\omega t ~{\bf i}
+\sin \theta \sin \omega t ~ {\bf j}+\cos \theta ~ {\bf k}]$ where $B_o$ is the
constant field strength. The scattering region is characterized by an energy 
level $\epsilon=\epsilon_o-q V_g$ which can be controlled by the gate 
voltage $V_g$ (see inset of Fig.1). We have only included the coupling between 
magnetic field and the spin degrees of freedom. It is however not difficult
to confirm that the orbital degrees of freedom do not contribute to the 
current in the presence of time varying magnetic field. This is because in 
the presence of magnetic field, the hopping matrix element between sites i 
and j, $t_{ij}$, in the tight binding description, will be modified by a 
phase factor $\exp[i\phi_{ij}]$ with $\phi_{ij}={\bf A}\cdot ({\bf r}_i - 
{\bf r}_j)$. But $\phi_{ij}$ due to our rotating magnetic field in the $x-y$ 
plane is simply zero, therefore the orbital parameter $t_{ij}$ is not
affected by the rotating field. The third term in Eq.(\ref{eq1}) denotes 
coupling between the scattering region and lead $\alpha$ with coupling 
matrix elements $T_{k\alpha}$. In the following we solve the transport 
properties (charge and spin currents) of the model in both adiabatic and 
non-adiabatic regimes using the standard Keldysh nonequilibrium Green's 
function (NEGF) technique\cite{jauho,wbg}.

{\bf Adiabatic regime.} Adiabatic regime is when the external 
parameter varies very slowly, {\it i.e.} in the $\omega\rightarrow 0$ limit. 
In this regime the charge with spin $\sigma$ transported from lead $\alpha$ 
in time interval $dt$ is given by\cite{wei}
\begin{equation}
dQ_{\alpha \sigma}(t) = q\int \frac{dE}{2\pi} (-\partial_E f) \left[ 
\Gamma_\alpha {\bf G}^r(t) {\bf \Delta} {\bf G}^a(t) \right]_{\sigma 
\sigma} dt
\label{charge}
\end{equation}
where ${\bf G}^r(t)$ is the retarded Green's function and 
${\bf G}^a(t)=[{\bf G}^r(t)]^{\dagger}$ is the advanced Green's function.
In the adiabatic limit,
\begin{equation}
{\bf G}^{r}(t)=\frac{1}{z}\left(
\begin{tabular}{cc}
$E-\epsilon_2$ & $\gamma e^{-i\omega t}$ \\ 
$\gamma e^{i\omega t}$ & $E-\epsilon_1$
\end{tabular}
\right)\ \ \ , 
\label{gr}
\end{equation}
where $z\equiv (E-\epsilon_1)(E-\epsilon_2)-\gamma^2$, $\epsilon_{1,2}
\equiv \epsilon \pm B_{0}\cos \theta -i\Gamma/2$, 
and $\Gamma=\sum_\alpha \Gamma_\alpha$ is the linewidth 
function. We will apply the wideband limit so that $\Gamma$ is independent
of energy. In Eq.(\ref{charge}), quantity ${\bf \Delta} \equiv dH'/dt$ where 
$H'$ is the $2\times 2$ matrix in spin space given by Eq.(\ref{hp}),
\begin{equation}
H'=\left(
\begin{tabular}{cc}
$0$ & $\gamma e^{-i\omega t}$ \\ 
$\gamma e^{i\omega t}$ & $0$
\end{tabular}
\right) \ \ .
\label{hp1}
\end{equation}

Using Eqs.(\ref{charge},\ref{hp1},\ref{gr}), the instantaneous electric current 
is found to be (Fermi energy and temperature are set to zero):
\begin{equation}
\frac{dQ_{\alpha \uparrow}}{dt} =-\frac{dQ_{\alpha \downarrow}}{dt}
=\frac{q\omega \Gamma_\alpha \Gamma \gamma^2}{(2\pi|\epsilon_1 \epsilon_2
-\gamma^2|^2)}\ \ .
\label{dqdt1}
\end{equation}
Several observations are in order for this result. First, spin flip 
mechanism (due to Eq.(\ref{hp1}))
in the scattering region is provided by the rotating magnetic field with
processes in which photons are absorbed and re-emitted. As a result, the 
{\it instantaneous} current is actually time-independent as Eq.(\ref{dqdt1}).
Second, the total instantaneous charge current
$dQ_{\alpha\uparrow}/dt+dQ_{\alpha\downarrow}/dt=0$ identically, {\it i.e.}
a rotating magnetic field does not pump a charge current.  Third, There is a 
nonzero spin current 
$I_{s\alpha} \equiv ds/dt=(1/q)dQ_{\alpha\uparrow}/dt$. 
From Eq.(\ref{dqdt1}), the spin current depends on field strength $B_o$, 
frequency $\omega$ of the rotating field, and more importantly on the energy
level positions $\epsilon_{1,2}$ which is modulated by the gate voltage.  It
is this modulation which provides the operation principle of our SFET.

The maximum spin current in the adiabatic regime is obtained by setting 
$\theta=\pi/2$ and $\Gamma_\alpha=\gamma=\Gamma/2$, we have
\begin{equation}
I_{s\alpha} = \frac{\omega}{4\pi} \frac{\Gamma^4/4}{\epsilon^4+\Gamma^4/4}
\ \ .
\label{spin}
\end{equation}
This lineshape---involving fourth power of the relevant quantities, is ideal 
for SFET operation: $I_{s\alpha}$ is very sensitive to the energy level 
position which is controlled by the gate voltage. For instance, at resonance 
$\epsilon=0$ the spin current reaches its maximum value $\omega/4\pi$. 
However, when $\epsilon$ is varied by $V_g$ to $10 (\Gamma/\sqrt{2})$,
the spin current is reduced by a factor of $10^4$. Since $I_s = s/\tau$ with 
$\tau=2\pi/\omega$ being the period of rotating magnetic field, we therefore 
conclude that at resonance, our device pumps out exactly one spin through the 
left or the right lead in one rotation. This quantization of pumped spin is 
substantially easier to realize than that of charge\cite{levinson,wang1} in a 
charge pump. It is easy to show that for a multi-probe system such as ours, 
the total spin pumped out of the device is two spin quanta. But if there is 
only one lead connected to the scattering region, the spin current is given by 
Eq.(\ref{spin}) multiplied by a factor of two: in this case our device
can be viewed as a nonmagnetic version of spin battery\cite{brataas2,foot}.

Why our device can pump out a DC spin current without a bias? As pointed 
out by Avron et al\cite{avron}, in a quantum parametric {\it charge} pump, 
the pumped charge per cycle is related to the Berry's phase\cite{berry}. This
physical picture can be easily generalized to the case of spin current
discussed in this work. In fact, using the spinor $|\Psi>= 
\left( \begin{array}{c} s_{11}\\s_{21} \end{array} \right)$ with
$s_{ij}$ the scattering matrix, the pumped 
charge can be obtained\cite{avron} from the definition of Berry's phase 
$\gamma = \int_0^\tau {\bar \gamma}(t) dt$ where ${\bar \gamma}(t)=
i<\Psi({\bf R}(t))|{\dot \Psi({\bf R}(t))}>$, ${\bf R}(t)$ label the 
slowly varying system parameters, and $\tau$ is the period of variation. 
Note that in the case of charge pumping, ${\bar \gamma}(t)$ corresponds to 
the instantaneous pumped charge. Setting $T_{k\alpha}=0$ in Eq.(\ref{eq1}), 
it is easy to verify\cite{ni} that ${\bar \gamma}(t)$ (instantaneous phase) 
is independent of time.

{\bf Non-adiabatic regime.} The electric and spin current beyond the adiabatic 
approximation can be calculated exactly using NEGF. It is convenient to define 
the particle current operator in spin space,
\begin{eqnarray}
\hat{J}_{\alpha ,\sigma \sigma ^{\prime }} &\equiv &\sum_{k}\frac{
d[C_{k\alpha \sigma }^{+}C_{k\alpha \sigma ^{\prime }}]}{dt} \label{j1} \\
&=&-i\sum_{k}[T_{k\alpha }C_{k\alpha \sigma }^{+}d_{\sigma ^{\prime
}}-T_{k\alpha }^{\ast }d_{\sigma }^{+}C_{k\alpha \sigma ^{\prime }}] 
\ . \nonumber
\end{eqnarray}
Then the electric current operator is $\hat{I}_{\alpha q}=q\sum_{\sigma}
\hat{J}_{\alpha ,\sigma \sigma}$
and the spin current operator is ${\bf I}_{s\alpha}=\sum_{\sigma 
\sigma ^{\prime }} \hat{J} _{\alpha ,\sigma 
\sigma ^{\prime }}~{\bf s}_{\sigma \sigma^{\prime }}$ 
where ${\bf s} = {\bf \sigma}/2$. From (\ref{j1}) we compute particle current 
\begin{eqnarray}
&&J_{\alpha \sigma \sigma ^{\prime }}(t) \equiv 
<\hat{J}_{\alpha ,\sigma \sigma^{\prime }}(t)> \nonumber \\
&=&-\sum_{k}[T_{k\alpha }G_{d\sigma ,k\alpha \sigma ^{\prime
}}^{<}(t,t)-T_{k\alpha }^{\ast }G_{k\alpha \sigma ^{\prime },d\sigma
}^{<}(t,t)] 
\label{particle_cur}
\end{eqnarray}
where the NEGFs are defined as
$G_{d\sigma ,k\alpha \sigma ^{\prime }}^{<}(t,t^{\prime })=i<C_{k\alpha
\sigma ^{\prime }}^{+}(t^{\prime })d_{\sigma }(t)>$,
$G_{k\alpha \sigma ,d\sigma ^{\prime }}^{<}(t,t^{\prime })=i<d_{\sigma
^{\prime }}^{+}(t^{\prime })C_{k\alpha \sigma }(t)>$. They are calculated by
the Keldysh equation ${\bf G}^< = {\bf G}^r \Sigma^< {\bf G}^a$ in standard
fashion\cite{jauho,wbg}. Therefore, the transport problem is reduced to the 
calculation of the retarded Green's function $G_{\sigma \sigma ^{\prime }}^{r}
(t,t^{\prime })$. 

In general, a perturbation theory is needed to solve a time-dependent problem. 
Fortunately, for the time-dependent Hamiltonian considered here, 
$G_{\sigma \sigma ^{\prime }}^{r}(t,t^{\prime })$ can be solved {\it exactly}
as follows. It is simple to obtain the retarded Green's function for the 
diagonal part (in spin space) of the Hamiltonian (\ref{eq1}),
\[
{\bf G}^{0r}(t-t^{\prime })=-i\theta (t-t^{\prime })\left(
\begin{tabular}{cc}
$e^{-i\epsilon _{1}(t-t^{\prime })}$ & $0$ \\ 
$0$ & $e^{-i\epsilon _{2}(t-t^{\prime })}$
\end{tabular}
\right) 
\]
The full Green's function of Hamiltonian (\ref{eq1}) is then calculated by
the Dyson equation in spin space, 
\begin{eqnarray}
{\bf G}^{r}(t,t^{\prime })&=&{\bf G}^{0r}(t-t^{\prime }) \nonumber \\
& & +\int dt_{x}{\bf G}^{0r}(t-t_{x})H^{\prime }(t_{x}){\bf G}^{0r}(t_{x}-t)
+\cdots  \nonumber
\end{eqnarray}
where $H'$ is given by Eq.(\ref{hp1}). After applying the double-time Fourier 
transform,
the Dyson equation can be summed up exactly to obtain the exact Green's 
function of the model (\ref{eq1}):
\[
G_{\sigma\sigma}^{r}(E,E^{\prime })=
\frac{ 2\pi \delta (E-E^{\prime })
G_{\sigma\sigma}^{0r}(E)}
{1-\gamma ^{2}g(E)}
\]
\[
G_{\sigma\bar{\sigma}}^{r}(E,E^{\prime })=2\pi 
\delta(E+\bar{\sigma}\omega -E^{\prime})
\frac{\gamma g(E)}{1-\gamma^2 g(E)}
\]
where $g(E)\equiv G_{\sigma\sigma}^{0r}(E)
G_{\bar{\sigma}\bar{\sigma}}^{0r}(E+\bar{\sigma}\omega)$, $\bar{\sigma}
=-\sigma$, and $\sigma = (\uparrow\downarrow) = \pm 1$.

Using these relations, it is straightforward to obtain particle current
from Eq.(\ref{particle_cur})
\begin{eqnarray}
J_{L\uparrow \uparrow } &=&-J_{L\downarrow \downarrow } 
=-\int \frac{dE}{2\pi }\Gamma_{L} \Gamma [f(E)-f(E_-)] \nonumber \\
&&\times \frac{\gamma ^{2}|G_{\uparrow \uparrow }^{0r}(E)|^2 |G_{\downarrow 
\downarrow }^{0r}(E_-)|^2}{|1-\gamma ^{2}G_{\uparrow \uparrow }^{0r}(E)
G_{\downarrow \downarrow }^{0r}(E_-)|^2} 
\label{particle1}
\end{eqnarray}
and $ J_{L\uparrow \downarrow }=0$, where $E_{-}\equiv E-\omega$. This
result\cite{foot1} allows us to conclude that the charge current is still 
identically zero while the spin current is given by
\begin{equation}
{\bf I}_{sL}=J_{L\uparrow \uparrow } {\bf k} 
\label{spin2}
\end{equation}
which is independent of time. These qualitative features are the same as
those of the adiabatic limit discussed above.  However, the non-adiabatic
result (\ref{particle1}) involves processes with energies $E\pm \omega$, as
shown by the arguments of the Green's functions.  This indicates that in the
general non-adiabatic situation, many {\it single} photon processes are
participating the operation of the SFET device.

{\bf Nanotube SFET.} We now apply the general principle of our SFET to a
(5,5) armchair single wall Carbon nanotube (CNT) with 200 unit cells
which is contacted by two electrodes and gated by a third probe, as shown 
in the inset of Fig.1. For simplicity, the CNT is modeled with the 
nearest-neighbor $\pi$-orbital tight-binding model with bond potential 
$V_{pp\pi}=-2.75$ eV for the carbon atoms. This model is known to give a 
reasonable, qualitative description of the electronic and transport 
properties of carbon nanotubes\cite{ref1}. Using Eq.(\ref{particle1}) the 
spin current flowing out of the CNT SFET in the adiabatic regime can be 
written as $I_s = \frac{\omega}{4\pi} T$ where
\begin{equation}
T=\frac{\Gamma^2 \gamma^2}
{(\epsilon^2+\Gamma^2/4-\gamma^2)^2+\Gamma^2 \gamma^2}\ \ .
\label{ns}
\end{equation}
Clearly, if $\gamma \leq \Gamma/2$, there is only one peak with $T \leq 1$. 
If $\gamma > \Gamma/2$, there are two peaks with $T=1$. We note that 
Eq.(\ref{ns}) is the same as that of the Andreev reflection coefficient
in the presence of superconducting lead (NS system). This can be understood 
as follows. For simplicity we assume a single probe connected to the
scattering region. Because of the rotating magnetic field, an incoming
spin-up electron (the electron in NS case) is flipped down and pumped out 
as a spin-down electron (the hole in NS case) which is analogous to the 
Andreev reflection. 
Fig.1 shows the spin current $I_s$ versus the gate voltage $V_g$ 
%
%
for different $\gamma$ with $\omega=0.01$ (corresponds to 86 MHz in
our units) and $\theta=88^o$. 
%
%
Here $\gamma=0.1$ corresponds to $B=0.06$ Tesla. Very similar results are 
obtained for other $\theta$. The resonant spin current transport is
clearly seen by which $I_s$ increases from practically zero to large values 
under the control of $V_g$. Fig.2 displays the spin current versus 
frequency using the nonadiabatic result Eq.(\ref{particle1}), with 
$\theta=50^o$, $\gamma=0.5$ and $V_g=0.0$.  The nonlinearity sets in at
about $\omega\sim 0.7$.
Finally
the inset of Fig.2 depicts spin current as a function of $\theta$ with 
$\omega=0.01$, $\gamma=0.5$, and $V_g=0.0$. The spin current is rather 
substantial for a wide range of angles.

In summary, we have demonstrated that a rotating magnetic field of uniform strength 
induces a spin current without a charge current, in coherent quantum conductors
without needing any magnetic material. The spin current is critically tunable
through the control of a resonance level in the system by an external gate voltage, 
thereby generating a field effect transistor operation. The physics behind this 
phenomenon is the spin-flip mechanism by the external field, but is ultimately
related to the quantum physics of the Berry's phase\cite{berry}. Because spin 
current can be detected using an idea proposed by Hirsch\cite{hirsch}, the
rotating frequency of the field needs not to be large, and the device structure
is quite typical, we believe the SFET proposed here should be experimentally
realizable.

\section*{Acknowledgments}
We gratefully acknowledge support by a RGC grant from the SAR Government of 
Hong Kong under grant number HKU 7091/01P, a CRCG grant from the
University of Hong Kong, and from NSERC of Canada and FCAR of Quebec 
(H.G).

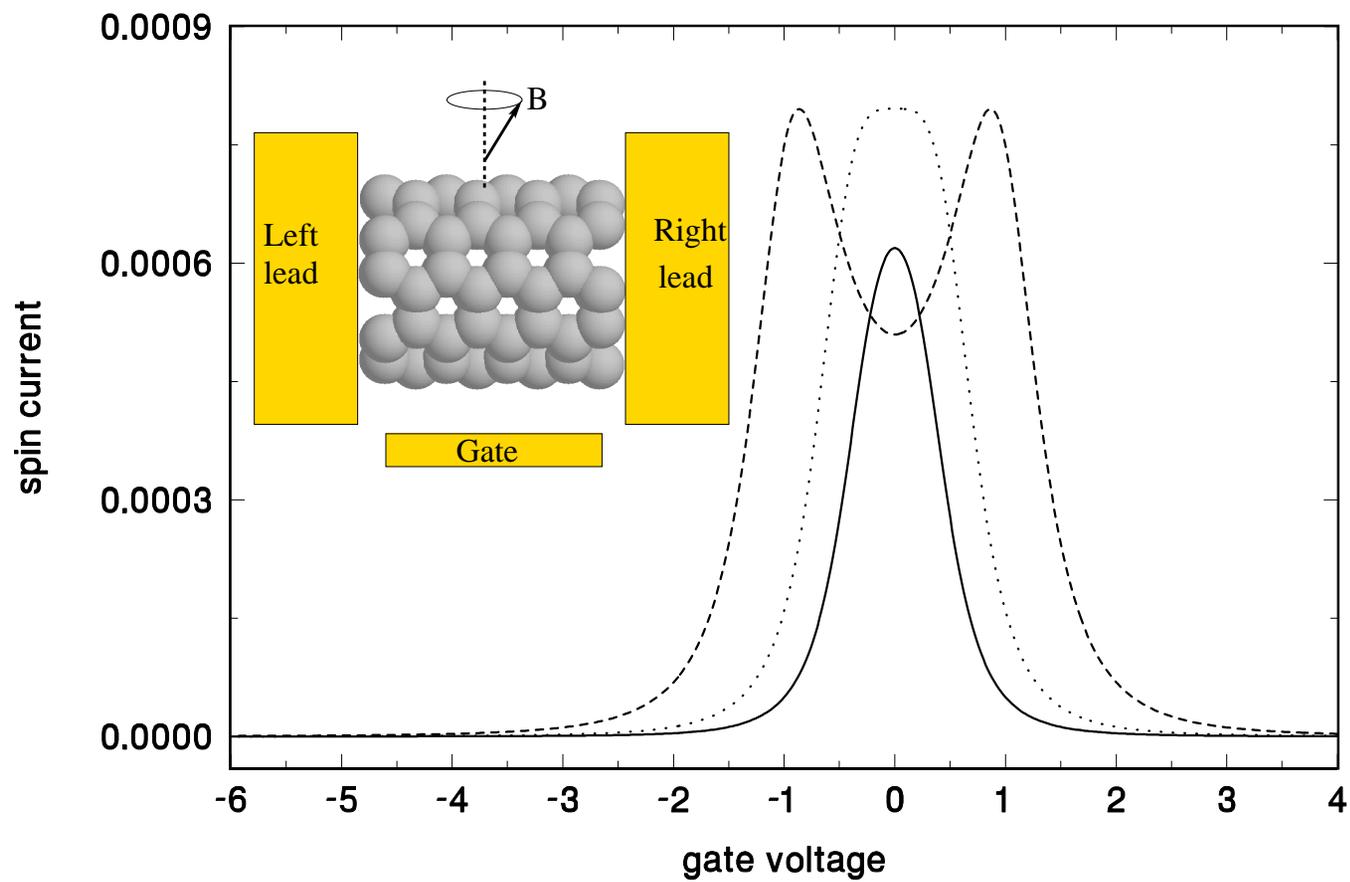
\begin{figure}
\caption{
The pumped spin current $I_s$ versus the gate voltage for different
$\gamma=0.3$ (solid line), $0.5$ (dotted line), and $1.0$ (dashed line). 
Inset: schematic plot of a nanotube SFET device. The energy unit in the 
calculation is $0.035$ meV.
}
\end{figure}

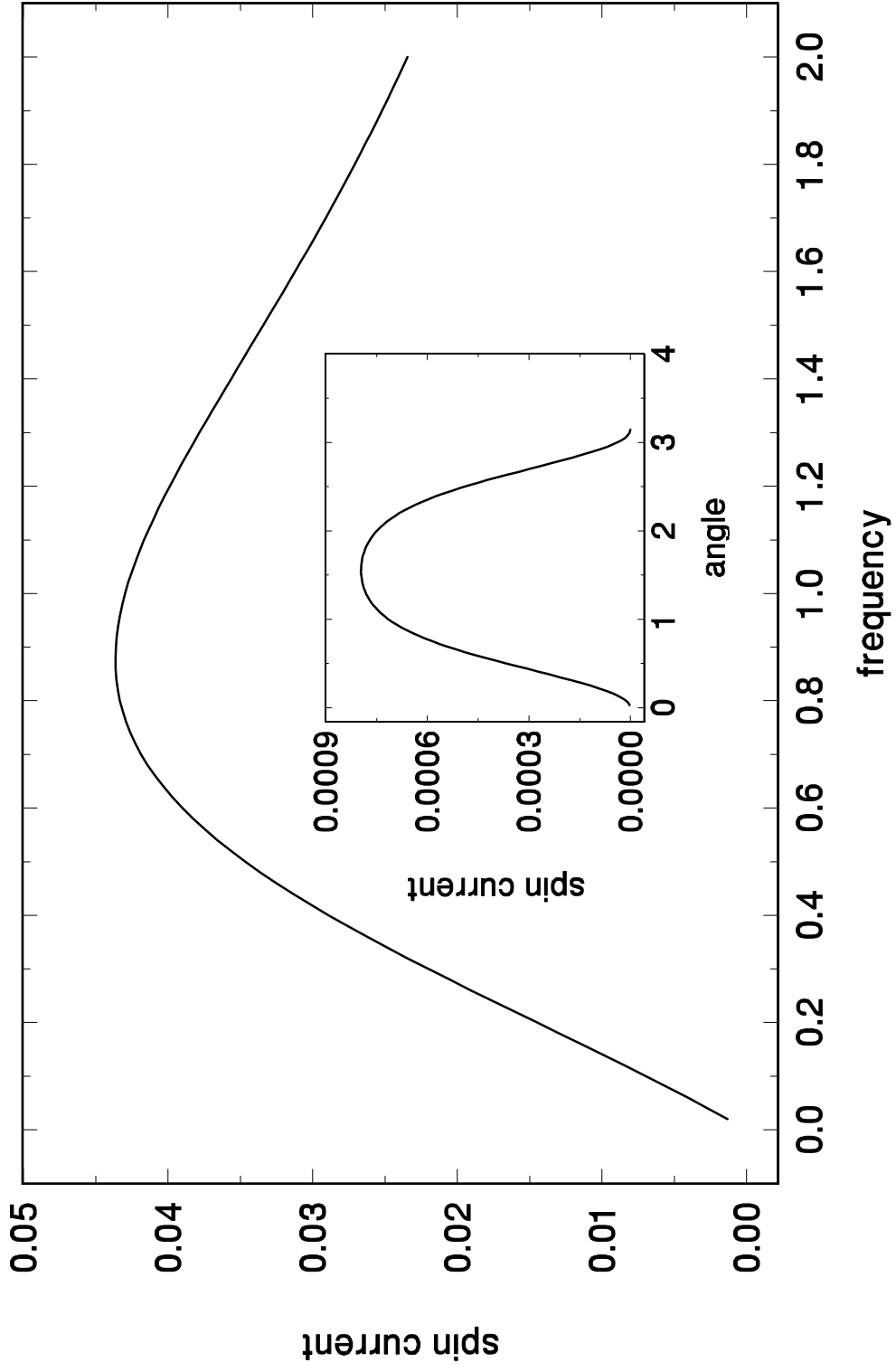
\begin{figure}
\caption{
$I_s$ versus frequency. Inset: $I_s$ versus the angle $\theta$.
}
\end{figure}


\begin{thebibliography}{00}
\bibitem{foot}
The spin battery concept of Ref.(\onlinecite{brataas2}) involves a
ferromagnetic---semiconductor interface and a rotating magnetic moment of
the ferromagnet. The system studied here can behave as a spin battery,
although there is no ferromagnetic material, if we contact the scattering
region by only one lead . From a device operation point of view, the scattering 
region has its own electronic structure (the energy levels) which induces 
resonance control of the spin current compared with a
ferromagnetic---semiconductor interface. Finally, our device is controlled
by an external field, rather than by internal magnetic moment. Therefore it
does not suffer from the damping effect of rotating magnetic moment that
consumes considerable amount of energy\cite{brataas1}.
\bibitem{foot1}
The integration in Eq.(\ref{particle1}) can be carried out exactly.
\end{thebibliography}
\end{document}